\documentstyle[aps,epsf]{revtex}
\begin{document}
\draft
\title{Nonextensive thermodynamic formalism for chaotic dynamical systems}
\author{{Ramandeep S. Johal $^1$} 
\thanks{e-mail: raman\%phys@puniv.chd.nic.in}
 {\it and} {Renuka Rai $^2$} \thanks{e-mail:rrai@mailcity.com}}
\address{{$^1$ \it Department of Physics, Panjab University,} \\
{\it Chandigarh -160014, India. }\\ 
 {$^2$\it Department of Chemistry, Panjab University,} \\
{\it Chandigarh -160014, India. }}
\date{\today}
\maketitle
\def\be{\begin{equation}}
\def\ee{\end{equation}}
\def\ba{\begin{eqnarray}}
\def\ea{\end{eqnarray}}

\begin{abstract}
A nonextensive thermostatic approach to chaotic dynamical 
systems is developed
by expressing generalized Tsallis distribution as escort
distribution. We explicitly show the thermodynamic limit
and also derive the Legendre Transform structure.
As an application, bit variance is calculated  
for  ergodic logistic map. 
Consistency of the formalism demands  
a relation between box size ($\epsilon$) and degree of nonextensivity,  
given as $(1-q)\sim -1/{\rm ln}\; \epsilon$.   
This relation is numerically verified for the case 
of bit variance as well as using basic definition of Tsallis entropy.
\end{abstract}
\section*{I. INTRODUCTION}
Techniques borrowed from Gibbs-Boltzmann or {\it extensive} thermodynamics
play important role in the characterization of complex
behaviour exhibited by dynamical systems. Analogous notions of
entropy, temperature, pressure and free energy can be applied
to quantify the fractal or multifractal attractors of chaotic
nonlinear mappings \cite{1}-\cite{3}. In fact, thermodynamic theory
of multifractals is applicable to any arbitrary probability distribution
irrespective of its mode of generation. A useful strategy here
is the following: for the given probability distribution 
$p_i={\rm exp}(-b_i)$, where $b_i$ is bit number, form
another set of normalized distributions $P_i = {p_i}^{\beta}/
\sum_{i}{p_i}^{\beta}$, called escort distributions  of order $\beta$.
For different $\beta$ values, the new distributions are better able to 
scan different features of the original distribution $p_i$.
Now in standard thermodynamics, maximizing the
Shannon-Boltzmann entropy $S=-\sum_{i} p_i {\rm ln}\;p_i$ under
the constraint of given mean value of internal energy,  
one obtains the canonical distribution,
$P_i = {\rm exp}(\Psi - \beta E_i)$, where $\Psi = -{\rm ln}\;Z(\beta)
 =- {\rm ln}\;(\sum_{i}{\rm exp}(-\beta E_i))$  
is equivalent to  Helmholtz free energy
and $Z(\beta)$ is  the partition function.
In chaos theory, the
 escort distribution is also sought in a 'generalized' canonical form
$P_i = {\rm exp}(\Psi - \beta b_i)$, where $\Psi = -{\rm ln}\;Z(\beta)$
and  $Z(\beta) = \sum_{i}{p_i}^{\beta} = \sum_{i}{\rm exp}(-\beta b_i)$. 
Thus in thermodynamic analogy,   positive
bit number $b_i$ is equivalent to energy $E_i$,
 $\beta$ plays the role of inverse temperature and $\Psi$ is
a generalized free energy. Knowing the
partition function is the first step to the study of analogous 
thermodynamic quantities and relations
in chaos theory. The analogy goes deeper
and interesting phenomena like phase transitions have been
highlighted in dynamical systems \cite{3}.

On the other hand, it has been realized in recent years  that the 
extensive formalism of thermodynamics 
fails to yield testable results (such as finite valued expressions 
for response functions of the various thermodynamic  quantities)
for systems with long-range interactions or
which evolve with
multifractal space-time constraints or have long term memory effects
\cite{4}.
To deal with such cases, a nonextensive formalism of 
statistical thermodynamics was proposed by Tsallis \cite{5}. 
 This formalism 
has many successful applications to its credit by now, and the importance
of nonextensive behaviour has been shown for diverse
kinds of systems: 
 L\'{e}vy anomalous diffusion \cite{5a,5b}, stellar polytropes \cite{5c}, 
pure-electron
plasma two dimensional turbulence \cite{5d,5e}, solar neutrinos \cite{5f},
inverse bremsstrahlung in plasma
 \cite{5g}, to name a few \cite{6}. Generally, this formalism is 
based on the proposal of a non-logarithmic
entropy 
\be
S_q(p) = \frac{1-\sum_{i}{p_i}^q}{q-1} = \sum_{i} [b_i] {p_i}^q.
\label{a}
\ee
Here, $[b_i]$ may just be considered as notation for the generalized
bit number within the nonextensive framework. For $q\to 1$,  
$[b_i] \to  b_i = -{\rm ln}\;p_i$ and 
Eq. (\ref{a}) gives the Shannon entropy. Tsallis entropy
is pseudo-additive (nonextensive) with respect to subsystems whose joint 
probabilities factorise:  if $p_{ij}^{I+II} = p_{i}^{I}
p_{j}^{II}$, then 
$S_q(p^{I+II}) = S_q(p^{I}) +  S_q(p^{II}) + (1-q)S_q(p^{I}) S_q(p^{II})   
$. Thus $(1-q)$ represents the deviation from extensivity.
Maximising  Tsallis entropy under the internal energy
constraint $U_q = \sum_{i} E_i {p_i}^q$ yields the equilibrium Tsallis
distribution \cite{7}
\be
p_i^{(eq)} = \frac{ \{1-(1-q)\beta E_i\}^{1/(1-q)}}{\sum_{i}\{1-(1-q)
\beta E_i\}^{1/(1-q)}},
\ee
where $\beta$ is the inverse temperature. For $q\to 1$,
we get back the canonical distribution. 

It is believed that Tsallis formalism is a natural frame to study
systems with fractal structures \cite{8}. Indeed the form of
Tsallis entropy was inspired from multifractal ideas \cite{5}.
 Also its connection with a 
scale-invariant thermostatistics has been conjectured \cite{9}. 
Simple relations exist between $q$ and characteristic exponents
of Levy flight distributions and anomalous diffusion \cite{10}.
Recently, a precise relation was proposed between $q$
and multifractal scaling properties of critical attractors 
of nonlinear one-dimensional maps \cite{11}.
For such cases, at the onset of chaos, 
the  exponential sensitivity to initial conditions is replaced
by power law sensitivity, and  
 generalizaton of Kolmogorov-Sinai entropy
has to be invoked on the lines of Tsallis formalism \cite{12}.

Realizing the clear signatures of nonextensivity in chaotic
dynamical systems, we feel a nonextensive thermodynamic 
approach needs to be developed for such systems.
In this paper, we develop  the thermostatistics of  such a formalism.
The plan of the paper is as follows: in section II, we present
our formalism by first writing the partition function, then 
seeking its thermodynamic limit. By generalizing the R\'{e}nyi
dimensions, we establish the Legendre Transform structure.
In section III, we present as an application, the calculation
of bit variance within nonextensive framework. In section IV,
we discuss the relation between box size (degree of coarse graining)
and degree of nonextensivity (1-q). Section V is devoted
to conclusions and outlook. 
\section*{II. GENERALIZED FORMALISM}
\subsection*{A. Basic ingredients}
  Our approach is to work with escort distributions
which are now sought not in  canonical form, but as 
 generalized Tsallis distribution. For 
this purpose, we consider a general, nonextensive statistics
 given by
\be
p_i = \{1-(1-q)\beta [b_i]\}^{1/(1-q)\beta}. \label{b}
\ee
For $\beta=1$, this gives the relation between $p_i$ and generalized
bit number $[b_i]$ in Tsallis statistics (see Eq. (\ref{a})). Also
for $q\to 1$, $p_i= {\rm exp}(-b_i)$, which corresponds to 
extensive statistics. Now the escort distribution for the given 
distribution of Eq. (\ref{b}) is
\be
P_i = \frac{{p_i}^{\beta}}{ \sum_{i}{p_i}^{\beta}}=
 \frac{ \{1-(1-q)\beta [b_i]\}^{1/(1-q)}}{\sum_{i}\{1-(1-q)
\beta [b_i]\}^{1/(1-q)}},
\label{b2}
\ee
which is  a generalized Tsallis distribution (see Eq. (2)), where now 
$[b_i]$ plays the role of energy $E_i$. The generalized
partition function, $Z_q(\beta)
=\sum_{i}{p_i}^{\beta} ={\sum_{i}\{1-(1-q) \beta [b_i]\}^{1/(1-q)}}$.
Again for $\beta=1$, escort distribution $P_i$ is identical to 
original distribution $p_i$. Note that in general, both the
original distribution and the escort distribution depend on $\beta$. 

Now an important feature of Tsallis
formalism is that it retains much of the formal structure 
of standard thermodynamics, including Legendre Transform structure
\cite{7}.  This is facilitated by a generalization of the logarithm
function as
\be
\Psi_q = - {\rm ln}_q\;Z_q(\beta) = -\frac{{Z_q}^{1-q}-1}{1-q}.
\label{c}
\ee
With the aid of such  modified logarithm (or conversely, the 
exponential function), we develop the nonextensive 
thermostatistics of multifractals.
Let us consider a certain nonlinear mapping over 
one-dimensional (for simplicity) phase space which has
been partitioned into boxes of equal size {$\epsilon$}. Let $p_i$
represent visiting frequency of the map for a box $i$.
Then we define  local crowding indices as
\be
 \alpha_i(\epsilon ,q)=\frac{{p_i}^{(1-q)\beta}-1}{(1-q)\beta
   {\rm ln}\;\epsilon}, 
\label{e}
\ee
which for $q\to 1$, can be written as $\alpha_i(\epsilon)=
\frac{{\rm ln}\;p_i}{{\rm ln}\;\epsilon}$. From Eqs. (\ref{b}) and  
 (\ref{e}), we may write
\be
[b_i]=\alpha_i(\epsilon ,q)V,  
 \label{f}
\ee
where $V =-{\rm ln}\;\epsilon$,
is the extensive volume parameter in  thermodynamic analogy.
\subsection*{B. Thermodynamic limit}
To  derive $\epsilon\to 0$  or  $V \to\infty$ limit of  
the generalized partition function,  
 we replace
the sum in   $Z_q = \sum_{i}{p_i}^{\beta}$  by an integral over $\alpha$ 
\cite{12a}, 
\be
Z_q=\int _{\alpha_{\rm min}}^{\alpha_{\rm max}}  d\alpha\; \gamma (\alpha )
\biggl(1-(1-q)\beta \alpha V\biggr)^{1/(1-q)},
\ee
where $\gamma (\alpha )d\alpha$ are the number of boxes with $\alpha_i$
in the interval $[\alpha$, $\alpha +d\alpha]$.
Now in extensive case, $\gamma (\alpha )$ is assumed to scale
as  $ \sim {\epsilon}^{-f(\alpha )} = {\rm exp}(f(\alpha ) V)$.
In the nonextensive case, this can be generalized as
$\gamma (\alpha )\sim (1+(1-q)f(\alpha ) V)^{1/(1-q)}$.
Thus we finally obtain
\be 
Z_q=\int_{\alpha_{\rm min}}^{\alpha_{\rm max}} d\alpha 
\biggl\{\biggl(1+(1-q)f(\alpha )V\biggr)
\biggl(1-(1-q)\beta\alpha V\biggr)\biggr\}^{1/(1-q)}.
\ee
It is useful to remember that the intergrand above is in a sense,  nonextensive
generalization of the function ${\rm exp}((f(\alpha )-\beta\alpha)V)$
in the extensive case. 
Now  applying the saddle point method for the case of large $V$, 
we may write the above integral as 
\be 
Z_q=\biggl\{\biggl(1+(1-q)f(\alpha )V\biggr)
\biggl(1-(1-q)\beta\alpha V\biggr)\biggr\}^{1/(1-q)}, 
 \label{k}
\ee
where $\alpha$ corresponds to the maximum value of the integrand 
for given $\beta$ value.
Then using Eq. (\ref{c}), we can write 
\be
\frac{\Psi_q}{V}=(\beta \alpha -f(\alpha ))-(1-q)\beta \alpha f(\alpha)V.
\label{siq}
\ee
Now in the limit $V\to\infty $,  the 
last term above diverges. To keep 
 the generalized free energy density finite, we assume that 
\be
(1-q)\beta \sim \frac{1}{V}. \label{bxq}
\ee
Then Eq. (\ref{siq}) is given by
\be
\lim_{\epsilon\to 0} \frac{\Psi_q}{V} \sim (\beta \alpha -f(\alpha ))
- \alpha f(\alpha).\ee
Note  that the first
term in parentheses is same as obtained for the extensive case. 
The second term arises due to the nonextensive
nature of the present formalism.  
Also it may be noted that if we assume volume parameter 
$V$ to be positive definite, then $(1-q)\beta \ge 0$.
\subsection*{C. Generalized R\'{e}nyi Dimensions}
Next, we propose to generalize the usual  R\'{e}nyi dimensions \cite{13}.
This is required to discuss Legendre Transform structure of the present 
formalism.  We define    
\be
D_q(\beta)=
\lim_{\epsilon\to 0}
\frac{1}{(1-\beta )V}{\rm ln\;}_q\biggl(\sum_{i} {p_i}^{\beta}\biggr).
\label{db1}
\ee
Again assuming the validity of Eq. (\ref{bxq}) for $\beta\ne 0$, 
in the $\epsilon\to 0$
limit, we get
\be
D_q(\beta) =  \frac{\beta }{1-\beta }\biggl(\biggl(\sum_{i} {p_i}^{\beta}\biggr)^{1-q}-1
\biggr). \label{p}
\ee         
For $\beta =0$, the last term of the Eq. (\ref{siq}) vanishes
and thus, Eq. (\ref{bxq}) cannot be inferred.
 Then using Eq. (\ref{db1}) and 
 $V = -{\rm ln}\; \epsilon$,  we get  
   $D_q(0) =-\lim_{\epsilon\to 0}   \frac{
 {\rm ln}_q\;N}{{\rm ln}\;\epsilon}$, where  $N$ is the number of nonempty
boxes. 
Also for $\beta =1$, Eq. (\ref{p}) gives $D_q(1) = -(1-q)\sum_{i} 
p_i{\rm ln}\;p_i$.
To see the behaviour for large $\beta$, we write from Eqs. (\ref{b})
and (\ref{f})
\be
{p_i}^{\beta} = \{1-(1-q)\beta \alpha_i V\}^{1/(1-q)}. 
\ee
Again in the limit of box size $\epsilon\to 0$, we assume the scaling
form Eq. (\ref{bxq}) and write
\be
{p_i}^{\beta} = \{1-\alpha_i \}^{1/(1-q)}.
\ee
For very large values of $\beta$, the above expression
is dominated by  minimum value of $\alpha$. Using this condition
in  Eq. (\ref{p}), we can write
\be
D_q(\beta) =  \frac{\beta }{\beta-1 }\alpha_{\rm min},
\ee
which for $\beta\to+\infty$ yields $D_q(\infty) = \alpha_{\rm min}$.
Similarly, for  $\beta\to-\infty$, one can show
$D_q(-\infty) = \alpha_{\rm max}$. These results are thus
identical to those obtained for the  usual R\'{e}nyi dimensions.
\subsection*{D. Legendre Transform structure}
Now from Eq. (\ref{p}), we write 
\be
Z_q=\sum_{i} p_i^\beta =\biggl\{1+\biggl(\frac{1-\beta }{\beta }\biggr)
D_q(\beta )\biggr\}^{1/(1-q)}.  \label{q}
\ee
Using Eq. (\ref{k}) with $\epsilon =0$,  and Eq. (\ref{q}), we get
\be
(\beta -1)D_q(\beta) =\beta \alpha -(1-\alpha )f(\alpha ),   \label{r}
\ee
which we choose to write as 
\be
\tau_q(\beta )=\beta \alpha -f_q(\alpha), \label{lts}
\ee
where we define $\tau_q(\beta )= (\beta -1)D_q(\beta) $ and
$f_q(\alpha)= (1-\alpha )f(\alpha )$.
 We find that
\be
\frac{\partial f_q(\alpha )}{\partial \alpha}=\beta,\quad
\frac{\partial \tau_q(\beta )}{{\partial \beta}}=\alpha.
\ee
We note   $\tau_q(\beta )$ being equivalent 
to the generalized free energy $\Psi_q$ (that already obeys Legendre
Transform structure), is a  concave function of its argument.
Then from Eq. (\ref{lts}), we infer that $f_q(\alpha)$ 
is also a concave function, implying that 
the  Legendre Transform structure is well defined.

\section*{III. GENERALIZED BIT VARIANCE: AN EXAMPLE}
An alternative method to characterize multifractal 
statistics is using bit moments and bit cumulants of
escort distributions \cite{zpb}.
Particularly, the second bit cumulant ($\Gamma _2$) 
measures the variance of bit number. In the following,
we derive expression for bit variance 
 within the generalized statistical framework
and apply it to ergodic logistic map. For nonextensive case, we write  
generalized second bit cumulant for escort distribution $P$ as 
\ba
{\Gamma_2}^{(q)}(P) = -{\beta}^2  
\frac{{\partial}^2\Psi_q}{{\partial}{\beta}^2}&=& 
q\biggl\{\biggl(\sum_{i}{p_i}^{\beta}\biggr)^{-q}\biggl(\sum_{i} 
{[b_i]}^2 {p_i}^{\beta(2q-1)}\biggr) \nonumber \\  
 & &  -\biggl(\sum_{i}{p_i}^{\beta}\biggr)^{-q-1}\biggl(\sum_{i}[b_i] 
{p_i}^{\beta q}\biggr)^2
\biggr\}.
\ea
The corresponding quantity for the original distribution ($\beta =1$)
is given by
\be
{\Gamma_2}^{(q)}(p) = -
\frac{{\partial}^2\Psi_q}{{\partial}{\beta}^2}{\bigg |}_{\beta =1} =
q\biggl\{\sum_{i} {[b_i]}^2 {p_i}^{(2q-1)}
-\biggl(\sum_{i} [b_i] {p_i}^{ q}\biggr)^2\biggr\}.
\label{gam}
\ee
Note that for $q\to 1$, we have ${\Gamma_2}(p)= \sum_{i} p_i ({\rm ln}\;
p_i)^2 - (\sum_{i} p_i {\rm ln}\;p_i)^2$. 
 For the case of ergodic maps, 
bit variance density  which is equivalent to heat capacity  is given as    
$C_2 = \langle ({\rm ln}\; \rho)^2\rangle -  \langle {\rm ln}\; \rho\rangle^2$
\cite{ssh},
where $\rho$ is the natural invariant density of the map. 
For the nonextensive case, we find  
\be
C_2^{(q)} = \frac{q}{(q-1)^2}( \langle {\rho}^{(2q-1)}\rangle
-\langle {\rho}^{ q} \rangle ^2).
\ee 
In Fig. 1, we plot $C_2$ and $C_2^{(q)}$ against the nonlinearity 
parameter $r$
for the logistic map, $x_{n+1} = rx_n(1-x_n)$. We find that for $0<q<1$,
the bit variance  (which represents size of fluctuations in bit number)
$C_2^{(q)}$ is larger than the  quantity $C_2$ of
extensive case. We present an interpretation of this
result in the next section. 

We also find quite an interesting feature {\it i.e.}
the general trend  of $C_2$  vs. $r$ curve, evaluated at smaller box size
 can be matched by $C_2^{(q)}$ evaluated at arbitrarily larger
box size and appropriate value of $q<1$. In Fig. 2, we
show two cases to corroborate this point. 
Quite remarkably, the two quantities show matching
for a wide range of the parameter $r$. 
To be precise, we note that decreasing the box size, lifts
the heat capacity curve $C_2$. Again, $q<1$ also results in a 
general increase in heat capacity $C_2^{(q)}$  (Fig. 1). 
Thus we can expect that heat capacity results with $q=1$
and smaller box size are reproduced with $q<1$ and larger
box size. This implies that a range of $q$ and $\epsilon$
values exist, which yield the same value for $C_2^{(q)}$.
In the following, we make use of this possibility.
\section*{IV. CONNECTION BETWEEN $\epsilon$ AND $(1-q)$} 
An important assumption in the above which makes the 
whole thermodynamic formalism consistent, is the
relation between box size $\epsilon$ and the degree
of nonextensivity $(1-q)$, as given by Eq. (\ref{bxq}).
Note that this relation is for an escort distribution
of order $\beta$. For the original distribution,
$\beta =1$ and we have $(1-q)\sim -\frac{1}{{\rm ln}\;\epsilon}$.    
In the following, we numerically check the valdity of this
scaling form. Again we consider the example of
section III. For exactness, we now focus on 
a fixed value of nonlinearity parameter $r =3.81$
in the chaotic region. The way we go about this
exercise is as follows: for a given value of $q$
and box size $\epsilon$, we calculate ${C_2}^{(q)}$. Then 
we vary both $q$ and $\epsilon$, so that 
we obtain same value of ${C_2}^{(q)}$ within
good approximation. Fig. 3 shows that a direct
proportionality in fact exists between $(1-q)$
and $-1/{\rm ln}\;\epsilon$, for very small $\epsilon$. 

One can look for the origin of the proposed relation
between  $(1-q)$ and $\epsilon$, in the definition
of Tsallis entropy itself. Thus consider a unit interval
divided into $W$ number of equal sized boxes. Assuming
equiprobability, we have $p_i =\frac{1}{W} =\epsilon$.
Then Tsallis entropy for this ditribution is given by
$S_q = \frac{1-{\epsilon}^{(q-1)}}{q-1}$.  
Again keeping $S_q$ fixed, we vary $q$ and $\epsilon$
and as shown in Fig. 4, find that the proposed scaling form  
$(1-q)\sim 1/V$ is justified for large $V$. 

Finally, we make a remark on the results of Fig. 1.
How do we interpret the higher values of ${C_2}^{(q)}$ 
over ${C_2}$ for $q<1$ in terms of effects
of nonextensivity ?
In the present context, we look at this issue from the
viewpoint of inherent  correlations due to nonextensivity.
It is known that if the  bit cumulants of subsystems, deviate
from additivity, this indicates the presence of correlations.
This happens naturally for the nonextensive case, where {\it e.g.}
the first bit cumulant, $\Gamma_{1}^{(q)}$ is identical
to Tsallis entropy and hence is nonadditive with respect
to statistically independent subsystems. Tsallis entropy
is superextensive for $0<q<1$, i.e. the  entropy of composite system 
is {\it larger} than the sum of entropies of individual subystems.
This implies that negative correlations are set up  for $0<q<1$.
We propose that 
the overall increase in bit variance for the nonextensive case above,
reflects these implicit correlations. 
In the thermodynamic limit of $V\to \infty$, the 
proportionality between $(1-q)$ and $1/V$ suggests
that $q\to 1-0$. Thus these correlations can be expected
to be suppressed in this limit.
Such correlations have also been studied in the context 
of classical ideal gas model within nonextensive
framework \cite{abe}.

\section*{V. CONCLUSION AND OUTLOOK}
In the present paper, we  presented a  
nonextensive generalization of thermostatistics of dynamical
systems. We have seen that in the thermodynamic limit, 
generalized free energy density is finite if we assume 
direct proportionality between $(1-q)$ and $1/V$. This
relation also plays important role in exhibiting
the Legendre Transform structure. We have numerically
shown the validity of this relation for generalized
bit variance and from the definition of Tsallis entropy 
itself.  
We know finite partitioning of the phase space is a fact of life
in the study of chaotic systems. That this feature can
be related to $(1-q)$, provides a new insight
into the nature of nonextensivity. 
Further development of the formalism
treating boxes with variable size, extension to
dynamical apsects involving time evolution 
will be extremely welcome. It can be hoped that these
possible developments will connect with the recent
work  \cite{11}, \cite{12} on low-dimensional dissipative systems and their
power law sensitivity to initial conditions at
bifurcation points or  onset of chaos.
\section*{VI. ACKNOWLEDGEMENTS}
RSJ would like to thank Professor Tsallis for useful
comments. RR acknowledges University Grants Commission, India
for the grant of Senior Research Fellowship.

\newpage
\begin{figure}
\caption{Heat capacity equivalent to bit variance is plotted against
control parameter $r$ of logistic map. Solid curve represents $C_2$
i.e. the extenive case, while dotted curve shows the behaviour of 
$C_{2}^{(q)}$.
$q$ value is set at $0.97$. Higher value of bit variance in the latter
case reflect the presence of nonextensive correlations.}  
\epsfbox{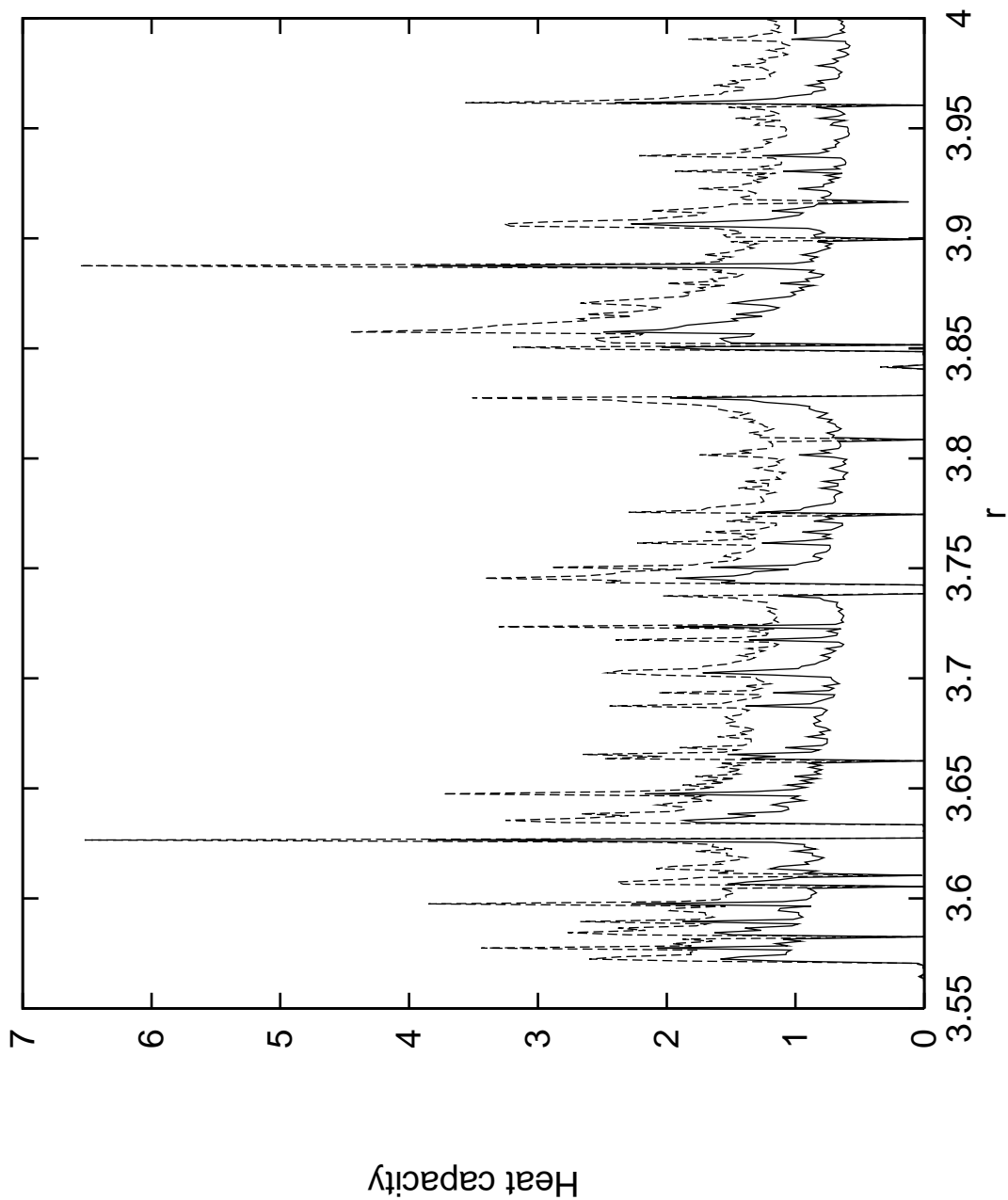}
\end{figure}
\clearpage
\begin{figure}
\caption{ $C_2$ (solid line) and $C_{2}^{(q)}$ (dashed line with symbols)
can be matched for a wide range of $r$ values: (a) $C_2$
with $2^{11}$ boxes. $C_{2}^{(q)}$ at $q =0.99$ with $2^{10}$
boxes. (b) $C_2$ with $2^{15}$ boxes. $C_{2}^{(q)}$ at $q =0.98$ 
with $2^{10}$ boxes. The plots show that decreasing $q$ from
unity in $C_{2}^{(q)}$, has the equivalent effect of 
decreasing the box size in $C_2$. Also note that $C_2$ increases
in general, on decreasing the box size.}
\epsfbox{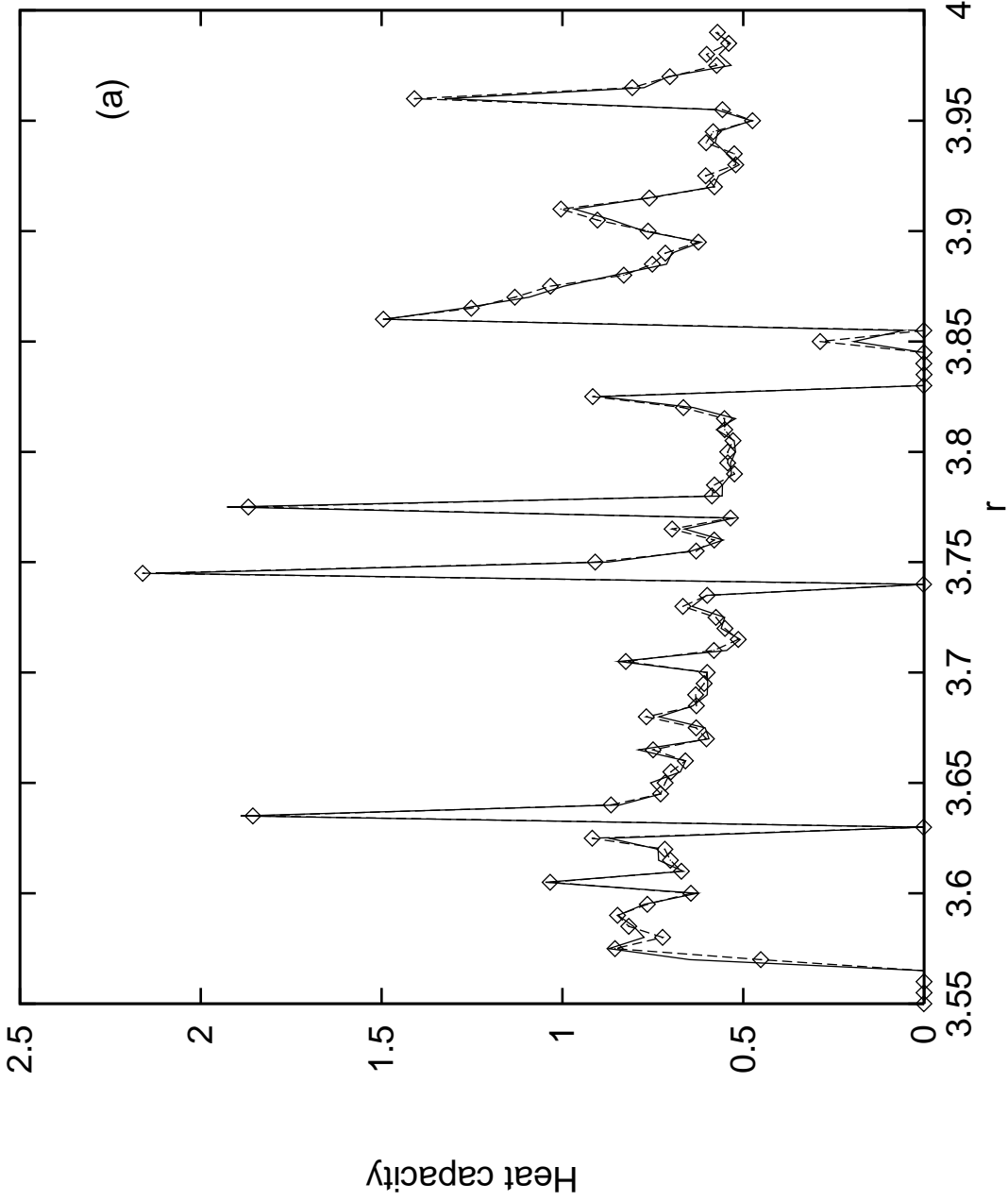}
\end{figure}
\clearpage
\begin{figure}
\epsfbox{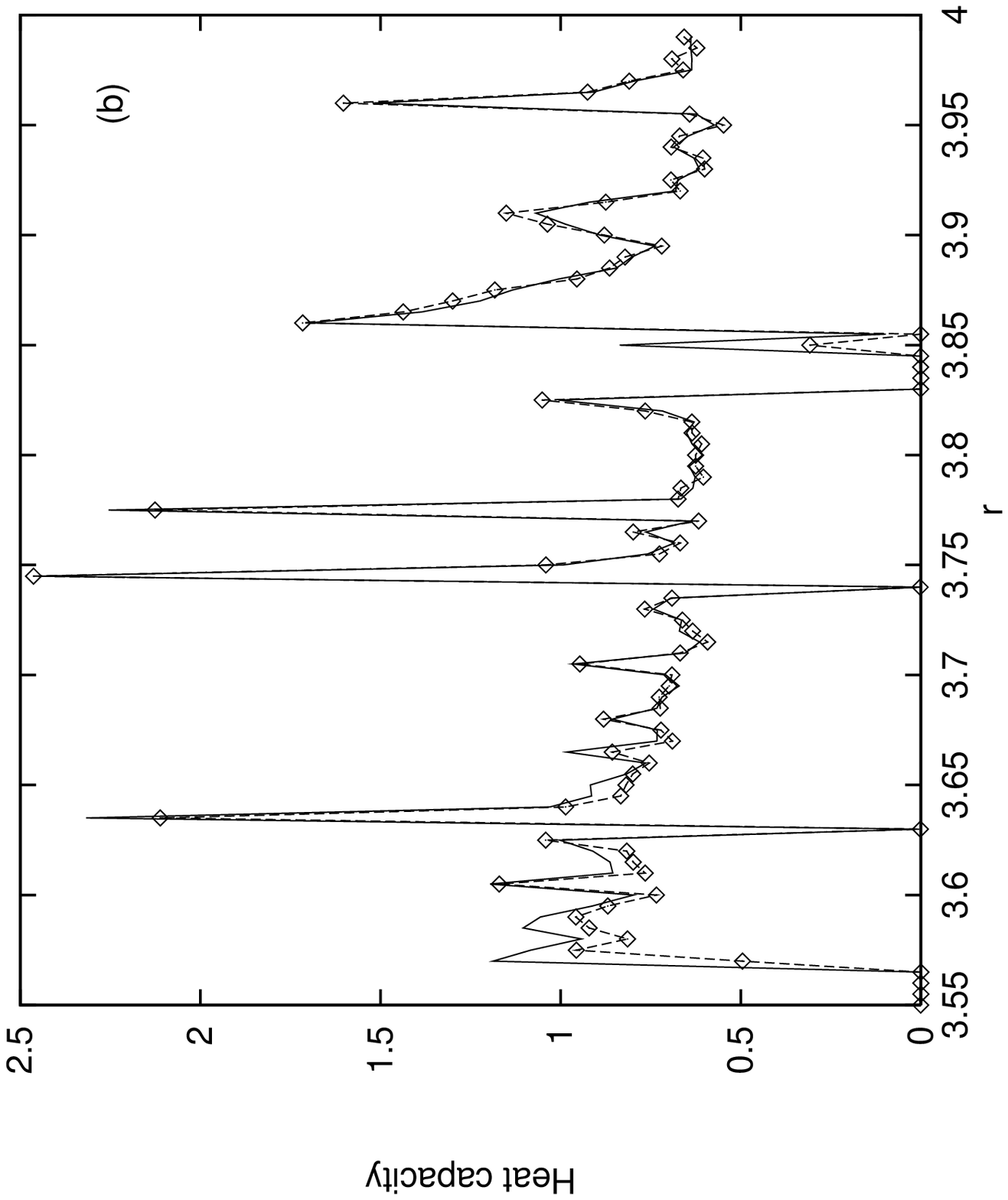}
\end{figure}
\clearpage
\begin{figure}
\caption{Plot between $(1-q)$ and $1/V$, keeping fixed $C_{2}^{(q)}$
value originally calculated at $r=3.81$, $q =0.9$ and box size
equal to 1/5000. The points can be fitted to straight line
$1.1x -0.032$.}
\epsfbox{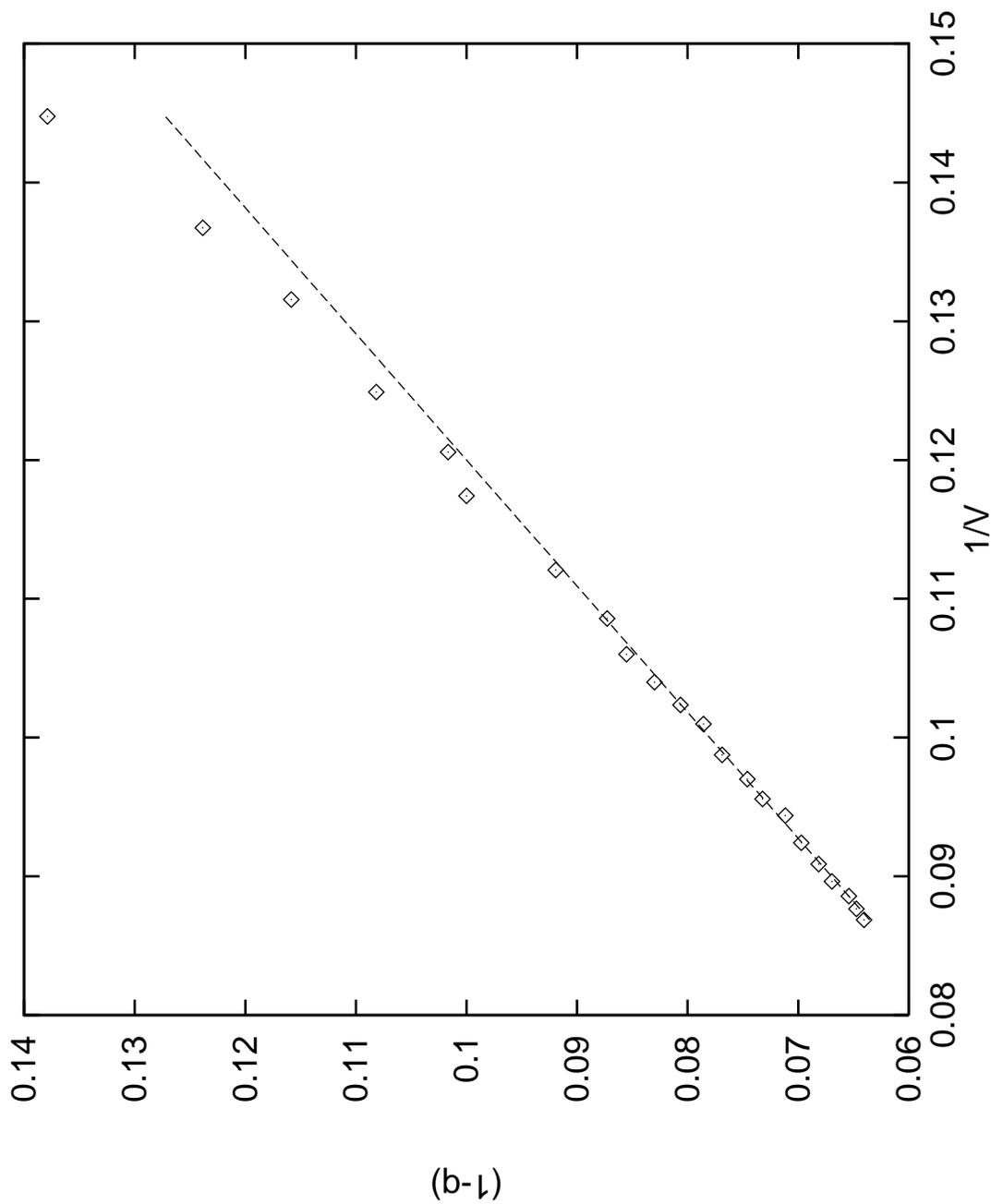}
\end{figure}
\clearpage
\begin{figure}
\caption{For equiprobablity, and $p_i =\epsilon$, the plot 
between $(1-q)$ and $1/V$, keeping fixed Tsallis entropy
originally calculated at $q =0.9$ and box size =  1/5000.
Straight line fit is given by $2.26x - 0.17$.}
\epsfbox{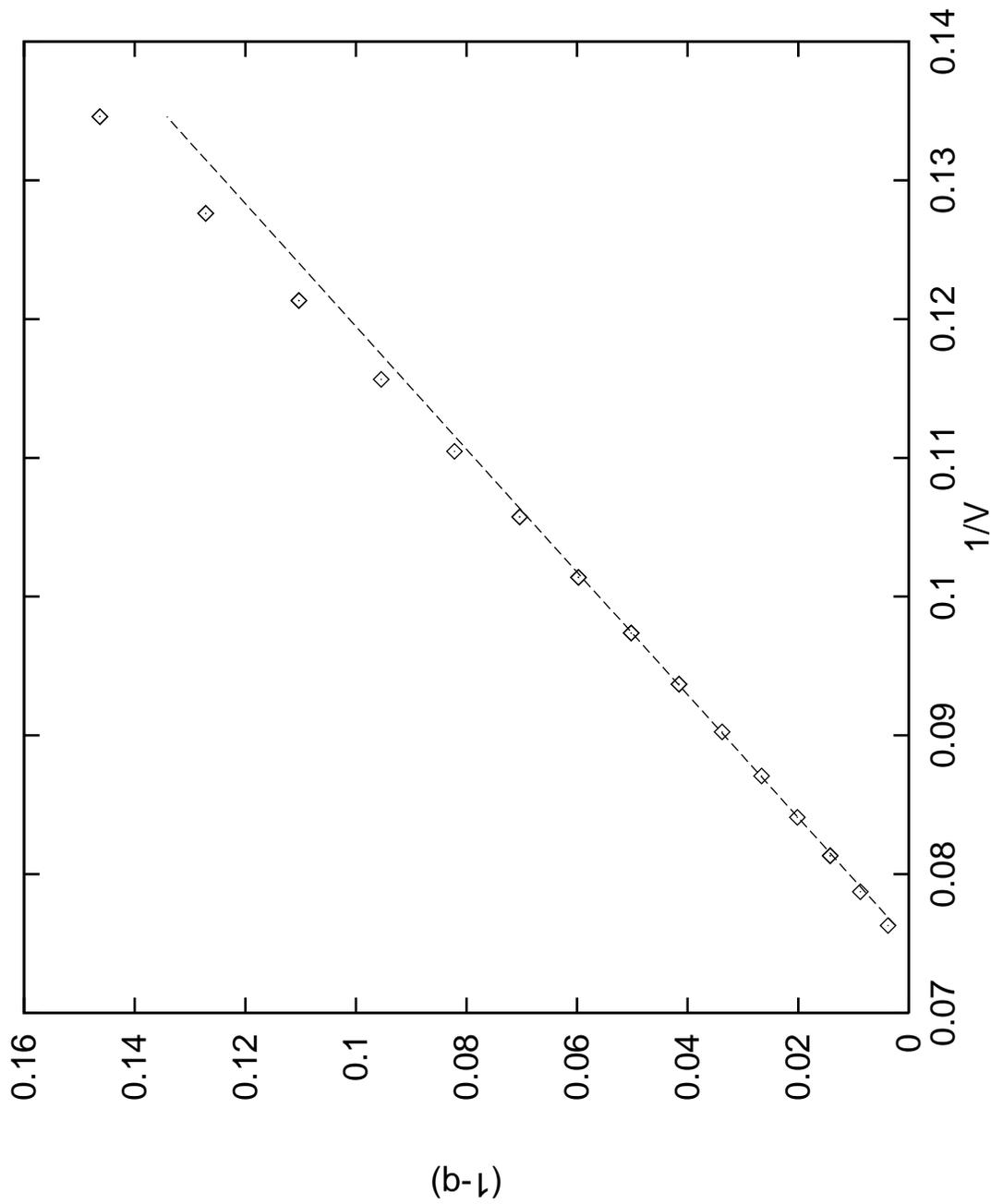}
\end{figure}
\end{document}